\documentclass[times,amsfonts,twocolumn,amsmath,amssymb,aps,prl]{revtex4-1}

\usepackage{color}
\usepackage{graphicx}
\usepackage{comment}
\usepackage{color}
\usepackage[colorlinks,urlcolor=blue,citecolor=blue,linkcolor=blue]{hyperref}
\usepackage{cleveref}
\usepackage{physics}
\usepackage{mathtools}
\usepackage{mathrsfs}
\usepackage{booktabs}
\usepackage{bm}

\usepackage{physics}
\usepackage{bm}
\usepackage[page,toc,titletoc,title]{appendix}
\usepackage{fixmath}
\usepackage{dsfont}
\hypersetup{
    colorlinks,
    citecolor=blue,
    filecolor=black,
    linkcolor=blue,
    urlcolor=blue
}
\usepackage{graphicx}
\usepackage{graphicx,accents}

\newcommand{\add}[1]{{\color{black}#1}}

\begin{document}

\title{Signatures of quantum chaos in an out-of-time-order tensor}

\author{Magdalini Zonnios}
\affiliation{School of Physics \& Astronomy, Monash University, Clayton, Victoria 3800, Australia}

\author{Jesper Levinsen}
\affiliation{School of Physics \& Astronomy, Monash University, Clayton, Victoria 3800, Australia}

\author{Meera M. Parish}
\affiliation{School of Physics \& Astronomy, Monash University, Clayton, Victoria 3800, Australia}

\author{Felix A. Pollock}
\affiliation{School of Physics \& Astronomy, Monash University, Clayton, Victoria 3800, Australia}

\author{Kavan Modi}
\email{kavan.modi@monash.edu}
\affiliation{School of Physics \& Astronomy, Monash University, Clayton, Victoria 3800, Australia}

\begin{abstract}
Motivated by the famous ink-drop experiment, where ink droplets are used to determine the chaoticity of a fluid, we propose an experimentally implementable method for measuring the scrambling capacity of quantum processes. Here, a system of interest interacts with a small quantum probe whose dynamical properties identify the chaoticity of the system. Specifically, we propose a fully quantum version of the out-of-time-order correlator (OTOC) -- which we term the out-of-time-order tensor (OTOT) -- whose correlations offer clear information theoretic meanings about the chaoticity of a process. We illustrate the utility of the OTOT as a signature of chaos using random unitary processes as well as in the quantum kicked rotor, where the chaoticity is tuneable.
\end{abstract}

\maketitle

The connections between quantum chaos and fast information scrambling have proven to be of practical and fundamental importance, particularly in understanding why isolated quantum systems thermalize~\cite{PhysRevA.43.2046, Nature2008Rigol, PhysRevLett.123.230606, PhysRevLett.125.180605, PhysRevE.95.062127}. Specifically, scrambling processes describe how local quantum information becomes lost in non-local degrees of freedom, i.e., due to entanglement~\cite{PhysRevA.99.062334, Landsman2019Nature,PhysRevA.43.2046}. Such processes look irreversible at the level of a local observable, analogous to what occurs in classically chaotic systems~\cite{strogatz2018nonlinear}.

This irreversibility is captured via the correlations of initially commuting local variables, before and after scrambling has occurred. This has been explored extensively via the decay of out-of-time-order correlators (OTOCs) $C(t)$ ~\cite{PhysRevA.99.052322,PhysRevA.99.051803,Sunderhauf2019, PhysRevLett.124.240505, PhysRevA.103.062214, PhysRevA.103.062214, PhysRevLett.126.030601} --- related to four-point temporal correlation functions of two initially commuting local Heisenberg operators $W$ and $V$~\cite{PhysRevLett.118.086801, 2017Roberts, PhysRevB.97.144304, PhysRevB.99.184202, PhysRevB.95.165136},
\begin{gather}\label{eq:f(t) otoc}
    F(t)=\expval{W_t V W^\dag_t V^\dag}_{\rho},
\end{gather}
via $C(t)=2(1-\Re[F(t)])$. Here, $\expval{\cdot}_{\rho} = \trace[\cdot \rho]$ with initial state $\rho$, and $W_t$ is the unitarily evolved operator $W$ in the Heisenberg picture at time $t$. For chaotic systems, $F(t)$ has the early-time exponential departure from unity: $1-\Re[F(t)]\approx e^{\lambda t}$ where $\lambda>0$~\cite{PhysRevLett.118.086801, Maldacena2016}. OTOCs thus act as an indicator of chaoticity by quantifying how quickly a local perturbation spreads into many-body correlations; the speed at which $W_t$ and $V$ fail to commute diagnoses the speed of information scrambling and the presence, or absence, of chaos~\cite{LewisSwan2019, Lantagne2019}.  

\begin{figure}
    \centering
    \includegraphics[scale=0.45]
    {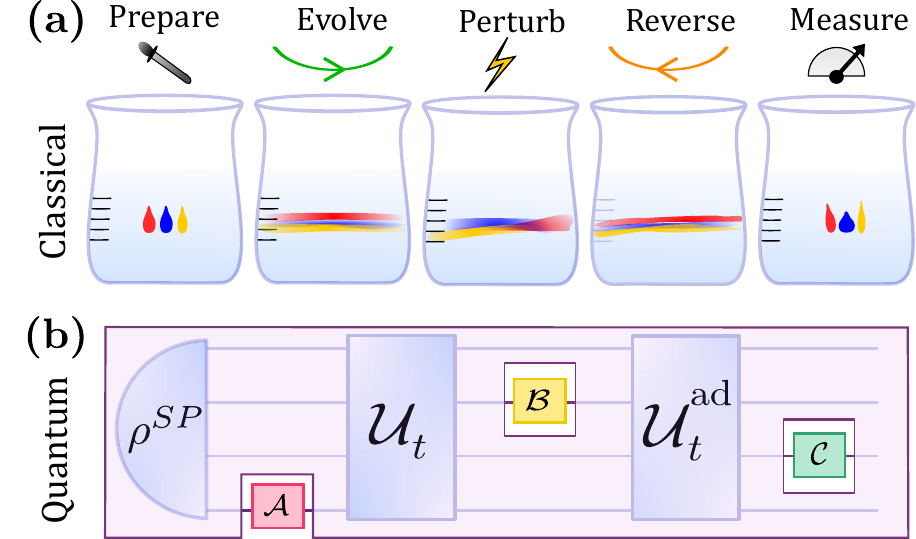}
    \caption{\textbf{(a) The ink-drop experiment} mixes ink in a viscous fluid by rotating the fluid, and subsequently demixed by rotating in the opposite direction. The indistinguishability of the initial and final states of the ink droplets, subject to a perturbation on the droplets or the fluid prior to the reverse rotation, quantifies the chaoticity of the fluid dynamics.
    \textbf{(b) The analog OTOC experiment} is as a higher order map (the object outlined in purple) that we refer to as an OTOT. The input-output correlations of the OTOT, i.e., between the preparation $\mathcal{A}$ and the measurement $\mathcal{C}$ of a small probe interacting with a large system, quantify the OTOC of this process.}
    \label{fig:OTOTcartoon}
\end{figure}

While the OTOC has been lauded as a way  
to capture many-body quantum chaos~\cite{PhysRevLett.121.124101, 2Blake2018, Blake2018, PhysRevB.98.205124, Cotler2020, PhysRevLett.126.030602}, it actually has strong overlap with the classical ink-drop experiment presented by David Bohm~\cite{BohmDavid2002Wati}. In this experiment, a cylindrical chamber is filled with a highly viscous fluid (e.g., corn syrup) and three droplets of ink are carefully placed in it. Due to the high viscosity of the fluid, the ink does not disperse as it would in water. Next, the fluid is rotated and the three colors mix. Remarkably, if the fluid is rotated backwards, the dispersed ink reforms into droplets resembling their initial shape (see Fig.~\ref{fig:OTOTcartoon}(a) or see~\footnote{\href{https://youtu.be/UpJ-kGII074}{https://youtu.be/UpJ-kGII074}.} for a real demonstration).

In this particular example, the dynamics of the fluid are regular, leading to a high similarity between the initial and reversed state of the ink droplets. On the other hand, if the viscous fluid were replaced with water, the rotation would cause chaotic (turbulent) dynamics such that the ink droplets could not easily be reversed to their initial state. In this sense, the similarity between the initial and reversed ink drops act as a probe of the chaoticity of the fluid as it is being rotated. Importantly, the steps of the ink-drop experiment are directly analogous to those of an experiment aiming to observe OTOCs. This is shown by expanding Eq.~\eqref{eq:f(t) otoc} in the Schr\"odinger picture~\cite{SM} to obtain the circuit shown in the bottom panel of Fig.~\ref{fig:OTOTcartoon}. Can we then adopt this idea to study quantum processes by interacting the system with a probe?

In this Letter, we derive an operator, the out-of-time-order tensor (OTOT), that simultaneously captures OTOCs with respect to all operators $V$ and $W$, i.e., a full quantum generalisation of the OTOC. The OTOT is a higher-order mapping from preparations to measurement outcomes, with an intermediate perturbation. It is shown to capture all possible correlations for an out-of-time-order process. The OTOT allows us to recast the decay of the OTOC in terms of information theoretic quantities such as the (conditional) quantum mutual information to indicate the chaoticity of the system. To illustrate this point, we study the OTOT for a spin-$\tfrac{1}{2}$ probe attached to a system $S$, where $S$ is either a random unitary process or a quantum kicked rotor, whose chaoticity is tunable.

\textit{Out-of-time-order tensor}.--- In classical physics the \textit{Kolmogorov-Sinai entropy}~\cite{sinai} relates temporal correlations of a process to its chaoticity. In the quantum realm, multitime correlations become higher-order maps, and their entropies can also be indicators for chaos~\cite{lindblad79, Karol, PhysRevLett.89.144101, cotler, PhysRevE.99.032213}. With this intuition, chaotic processes should have less correlations between an initial and time reversed probe, while regular processes should retain correlations faithfully throughout an out-of-time-order process. The complexity and chaoticity of a process in a system (e.g., the fluid) can hence be quantified by measuring the information of a coupled probe (e.g., the ink droplets). We now construct a fully quantum representation for the OTOC, i.e., a tensor capturing all facets of the out-of-time-order process. 

\begin{figure}
    \centering
    \includegraphics[scale=0.5]{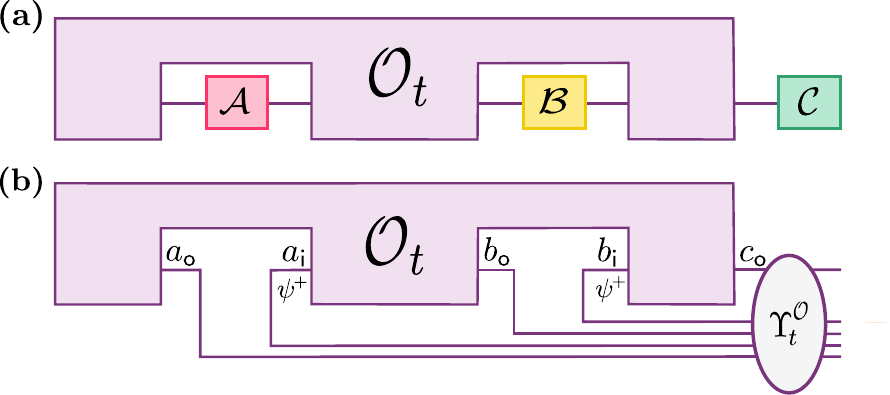}
    \caption{\add{\textbf{(a) OTOC from OTOT} obtained by contracting OTOT with superoperators $\mathcal{A}$, $\mathcal{B}$, $\mathcal{C}$ and tracing over the output, as per
    Eq.~\eqref{eq:the OTOT}. This reduces to the conventional OTOC when $\mathcal{A}[\rho]=V^\dag\rho\mathds{1}$, $\mathcal{B}[\rho]=W^\dag\rho W$, and $\mathcal{C}[\rho]=\mbox{tr}[\mathds{1}\rho V]$ ~\cite{SM}. This is equivalent to contracting the Choi state of the OTOT with the Choi states of superoperators $\mathcal{A}$, $\mathcal{B}$, $\mathcal{C}$. \textbf{(b) Choi state of OTOT}, obtained by inserting half of a maximally entangled state $\psi^+$ into each input of the OTOT. Each of the five wires of the OTOT represent the Hilbert space for different times in the process. Thus, the OTOT is a map on superoperators $\mathcal{A}$, $\mathcal{B}$, $\mathcal{C}$, which is commonly said to be a higher order map for the out-of-time-order process depicted in shaded purple in  Fig.~\ref{fig:OTOTcartoon}(b).}}
    \label{fig:ChoiState}
\end{figure}

We begin by considering the action of a set of operations on a small probe $P$, interacting with $S$. The probe is subject to a preparation $\mathcal{A}$ at the initial time, after which it evolves with $S$ under the unitary map $\mathcal{U}_{t}[\rho] := U_{t} \rho U^\dag_{t}$ for a time $t$. Next, a local perturbation $\mathcal{B}$ is applied either to $P$ or $S$. Finally, $SP$ is evolved back to the initial time with the adjoint unitary map $\mathcal{U}_{t}^{\text{ad}}[\rho] := U^\dag_{t} \rho U_{t}$, whereupon $P$ is measured with a measurement $\mathcal{C}$. This process, depicted in Fig.~\ref{fig:ChoiState}, defines the out-of-time-order map $\mathcal{O}_t$
\begin{align}\label{eq:the OTOT}
\begin{split}
    \mathcal{O}_t\left[\mathcal{C},\mathcal{B},\mathcal{A}\right] :=& \trace\left(\mathcal{C}\circ\mathcal{U}^{\text{ad}}_{t} \circ \mathcal{B} \circ\mathcal{U}_{t}\circ \mathcal{A}\left[\rho\right]\right)\\
    =& \trace[\Upsilon_t^\mathcal{O} (\hat\Upsilon_\mathcal{A} \otimes \hat\Upsilon_\mathcal{B} \otimes \hat\Upsilon_\mathcal{C})^T],
\end{split}
\end{align}
where $\mathcal{A}$, $\mathcal{B}$, and $\mathcal{C}$ are superoperators acting on density matrices with composition $\mathcal{A}\circ \mathcal{B}[\rho]:= \mathcal{A}[\mathcal{B}[\rho]]$. \add{For technical details on superoperators see, for example, Refs.~\cite{2017Milz, PRXQuantum.2.030201}.}

$\mathcal{O}_t$ acts on quantum maps $\mathcal{A}, \mathcal{B}, \mathcal{C}$, and thus is said to be an higher-order map~\cite{PhysRevLett.101.060401, PhysRevA.80.022339, Oreshkov2012} \add{that must possess non-Markovian quantum correlations~\cite{PollockPRL}, as we show below.} Rather than working with an abstract map, it is often convenient to work with its matrix representation $\Upsilon^\mathcal{O}_t$, which is the \textit{out-of-time-order tensor} (OTOT). The OTOT is given in the second line of Eq.~\eqref{eq:the OTOT}, obtained by means of the Choi-Jamio{\l}kowski  isomorphism~\cite{JAMIOLKOWSKI1972275, CHOI1975285}, which translates a quantum map, representing a process, to a many-body quantum state~\cite{2017Milz}. The isomorphism is shown in the bottom panel of Fig.~\ref{fig:ChoiState}, where each line from $\mathcal{O}_t$ to $\Upsilon_t^\mathcal{O}$ is an operator on space of $P$~\cite{SM}. The resultant OTOT is a density matrix satisfying ${\Upsilon_t^\mathcal{O}}^\dag = \Upsilon_t^\mathcal{O}$, $\Upsilon_t^\mathcal{O} \ge 0$, $\trace[\Upsilon_t^\mathcal{O}]=1$. By contrast, the Choi states $\hat\Upsilon_X$ of the superoperators $X\in \{\mathcal{A}, \mathcal{B}, \mathcal{C}\}$ play the role of observables and the hat denotes that they are not normalised; for a trace-preserving operation $\trace[\hat\Upsilon_X]$ equals the dimension of the Hilbert space. Hence, Eq.~\eqref{eq:the OTOT} has the interpretation as the spatio-temporal version of the Born rule~\cite{Shrapnel_2018}.

The importance of the OTOT lies in the fact that it is a full quantum embodiment of all possible OTOCs~\cite{1367-2630-18-6-063032,2018Pollock,  cotler}\add{, including $2k$-OTOCs~\cite{2017Roberts}}. Moreover, unlike the OTOC which looks at correlations between two \textit{fixed} observables, the OTOT provides a mapping from \textit{any} state preparation to measurement outcomes. The total correlations of the OTOT thus generalize the notion of an OTOC by providing a measure of the scrambling capacity of the process, independent of either preparation or measurement. The OTOT reduces to the conventional OTOC when $\mathcal{A}[\rho]= \mathcal{C}^{\text{ad}} [\rho] := V^{\dag}\rho\mathds{1}$ and $\mathcal{B}[\rho] := W^{\dag} \rho W$, i.e., we recover Eq.~\eqref{eq:f(t) otoc}, with $F(t) = \mathcal{O}_t[\mathcal{C}, \mathcal{B}, \mathcal{A}]$~\footnote{This particular choice of maps are not completely positive (CP), but could still be realised~\cite{PhysRevA.94.040302}.} --- see Supplemental Materials~\cite{SM}.

\textit{Information in OTOT}.--- Signatures of chaos can be seen in the correlations between subparts of the OTOT. (Henceforth, we assume an uncorrelated initial state $SP$ and drop $a_\mathsf{o}$ since, in practice, one would introduce a probe that is independent of the system~\cite{modiscirep, mi2021information}.) We now adopt the quantum mutual information (QMI) and conditional quantum mutual information (CQMI) to quantify bipartite and tripartite correlation, respectively. The former is defined as $I(x:y) := S(x) + S(y) - S(xy)$, where $S(xy):=-\tr[\Upsilon^{xy} \log \Upsilon^{xy}]$ is the von Neumann entropy of the density matrix $\Upsilon^{xy}$ and $\Upsilon^{x(y)} = \tr_{y(x)}[\Upsilon^{xy}]$. The CQMI $I_t(x:y|z)$ is defined by replacing each entropy in the last equation by a conditional entropy $S(xy|z):=s(xyz)-s(z)$. The QMI in $a_\mathsf{i} \to b_\mathsf{o} \ (b_\mathsf{i} \to c_\mathsf{o})$ quantifies the capacity of this channel to transmit information of the input $a_\mathsf{i} \ (b_\mathsf{i})$ to the output $b_\mathsf{o} \ (c_\mathsf{o})$. The correlations between $a_\mathsf{i} b_\mathsf{o}$ and $b_\mathsf{i} c_\mathsf{o}$ quantify how much the latter channel depends on the former. These are also known as non-Markovian correlations, and a process is Markovian if and only if these correlations are vanishing~\cite{PollockPRL}.

\begin{figure}
   \includegraphics[scale=.57]{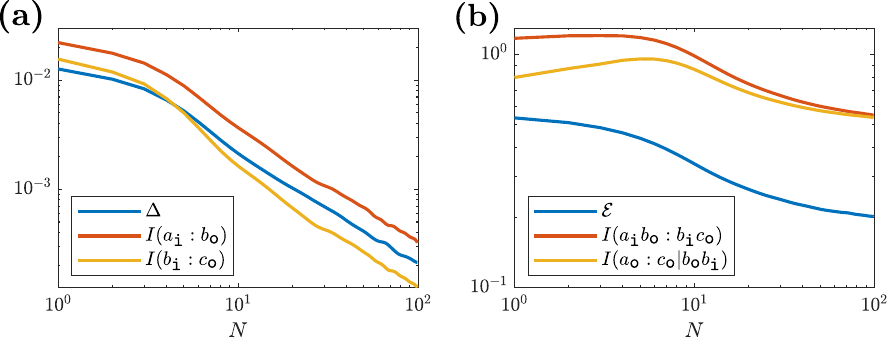}
    \caption{Information scrambling in the case of random unitary interactions between a spin-$\frac12$ probe and an $N$-dimensional system. \textbf{(a) Mutual information in the channels} $a_\mathsf{i} \to b_\mathsf{o}$ and $b_\mathsf{i} \to c_\mathsf{o}$ as in Fig.~\ref{fig:ChoiState}. 
    We also plot $\Delta$, given in Eq.~\eqref{eq:Delta}, whose fast decay is a result of the probe's strong sensitivity to chaos. \textbf{(b) Non-Markovian correlations} in terms of the quantum entanglement ($\mathcal{E}$(N)), the QMI between $a_\mathsf{i} b_\mathsf{o}$ and $b_\mathsf{i} c_\mathsf{o}$, and the CQMI between $a_\mathsf{o}$ and $c_\mathsf{o}$ given $b_\mathsf{o} b_\mathsf{i}$. The entanglement is given by the negativity $\mathcal{E} := \frac{1}{2}(\|\Upsilon_X^{T}\|-1)$ with $\|\cdot\|$ denoting the trace norm.}
    \label{fig:random matrix}
\end{figure}

In chaotic systems, we anticipate information scrambling and thus the forward $a_\mathsf{i} \to b_\mathsf{o}$ and backwards $b_\mathsf{i} \to c_\mathsf{o}$ channels will be highly noisy leading to low information transfer and consequently low two-time correlation. However, as stated above, the OTOT must be non-Markovian and the coherent coupling between these channels must possess non-trivial four-time (or non-Markovian) correlations. That is, we expect that information lost by the probe to the system upon forward chaotic evolution must return back to the probe subject to reversing the dynamics, even if an intermediate perturbation is made. In Fig.~\ref{fig:random matrix}, we show that, unlike traditional OTOCs, the OTOT is capable of identifying such non-Markovian correlations -- some of which are genuinely quantum, i.e., entangled -- and hence distinguishing between decoherence due to noisy Markovian dynamics and information scrambling due to chaotic non-Markovian dynamics.

We plot the QMI for various partitions of the Choi state for Haar random unitary interactions (averaged over 50 iterations) between a spin-$\tfrac{1}{2}$ probe and an $N$-dimensional system. That is, the role of $\mathcal{U}$ in Fig.~\ref{fig:OTOTcartoon}(a) and Eq.~\eqref{eq:the OTOT} is taken to be from a uniform distribution of unitaries. It is known that such unitaries are in fact highly entangling and generate ergodic (chaotic) dynamics.
The left panel shows how the correlations in the channels feature a powerlaw decay, corresponding to exponential decay in the `number of qubits' $\log_2(N)$. The right panel instead shows the correlations between the channels $a_\mathsf{i} \to b_\mathsf{o}$ and $b_\mathsf{i} \to c_\mathsf{o}$ (including entanglement), decay to a finite value. The fast decaying local and slowly decaying global correlations indicate increasing scrambling of the process with $N$. This is in contrast to decoherence where both local \textit{and} global correlations decay rapidly. For random circuits, non-Markovianity may also decay rapidly~\cite{arXiv:2004.07620, FigueroaRomero2019almostmarkovian}, but this is not the case for the OTOT because the forward and reversed dynamics are highly correlated.

A direct consequence of the above discrepancy is that the channel $a_\mathsf{i} \to c_\mathsf{o}$ can behave radically differently depending on the choice of operation $\mathcal{B}$, i.e., the effect of a small perturbation can lead to amplified effects over time --- the so-called `butterfly effect'~\cite{ALEINER2016378, Robert2015}.  Here, the perturbation is applied only to a single qubit of a many-qubit system. The Choi state, conditioned on a butterfly operation $\mathcal{B}$, is $\Upsilon^{a_\mathsf{i} c_\mathsf{o}|\mathcal{B}}_t := \tr_{b_\mathsf{o}b_\mathsf{i}}[\Upsilon^{\mathcal{O}}_t \hat\Upsilon_\mathcal{B}^{\mathrm{T}}]$~\cite{2017Milz}. We define the ratio of the minimum to maximum CQMI,
\begin{gather}\label{eq:Delta}
    \Delta := \frac{\min_{\mathcal{B}} I(a_\mathsf{i}:c_\mathsf{o}|\mathcal{B})}{\max_{\mathcal{B}} I(a_\mathsf{i}:c_\mathsf{o}|\mathcal{B})},
\end{gather}
to quantify the sensitivity to butterfly operations. The recent results by the Google group~\cite{mi2021information} study the numerator of $\Delta$ as a function of circuit depth.

For concreteness, we confine ourselves to unitary perturbations $\mathcal{B}$, which ensures the decay in correlations are purely due to scrambling. Without loss of generality, we take $\mathcal{B} [\rho] = e^{-i \phi \sigma_z} \rho e^{i \phi \sigma_z}$. This channel has the form~\cite{SM}
\begin{align}\label{eq: Choi for two-level}
    \Upsilon^{a_\mathsf{i}c_\mathsf{o}|\mathcal{B}}_t =& \cos^2(\phi) \psi^+ +\sin^2(\phi) \tr_S[Z_t \psi^+ \otimes \rho_S  Z_t^\dag] \\\notag
    &+ i \cos(\phi) \sin(\phi) \tr_S[Z_t \psi^+ \otimes \rho_S - \psi^+ \otimes \rho_S  Z_t^\dag],
\end{align}
where $Z_t := U_t \sigma_z \otimes \openone U_t^\dag$, $\rho_S$ is the initial state of the system and $\psi^+:= \dyad{\psi^+}$, where $\ket{\psi^+} := \tfrac{1}{\sqrt{2}} (\ket{00}+\ket{11})$ is the Bell state of the probe together with an ancillary two-level system~\cite{SM}. 

In the absence of a butterfly perturbation ($\phi=0$), all correlations trivially survive the process: $\max_{\mathcal{B}} I(a_\mathsf{i}: c_\mathsf{o}|\mathcal{B}) \to 2$. Conversely, the minimum of $I(a_\mathsf{i} : c_\mathsf{o}|\mathcal{B})$ is at $\phi = \pi/2$, corresponding to $\Upsilon^{a_\mathsf{i} c_\mathsf{o}|\mathcal{B}}_t = \tr_S[Z_t \psi^+ \otimes \rho_S  Z_t^\dag]$, which for highly entangling processes displays no correlations between $a_\mathsf{i}$ and $c_\mathsf{o}$. Thus, the decay of $\Delta$ depends on the chaoticity of the process. To see this, we now consider a physically realisable model where the chaoticity is tunable.

\textit{Example: Quantum kicked rotor with a spin-$\tfrac{1}{2}$ probe}.---We compute the correlations in an OTOT as a measure of the scrambling capacity of the quantum kicked rotor (QKR). This is a well-known model in both the classical and quantum chaos literatures with tunable chaoticity, as well as a clear correspondence between classical and quantum chaos. This makes it an ideal candidate for this investigation.

The classical kicked rotor (CKR) is paradigmatic to study transitions from integrability to chaos~\cite{NolteDavidD2019HC}, especially due to its simple dynamics~\cite{10.2307/27854538}. Specifically, the CKR describes a freely rotating pendulum which is subject to periodic kicks with period $\tau$. Its dynamics are governed by the (dimensionless) Hamiltonian~\cite{CHIRIKOV1979263}, 
\begin{gather}
\label{eq:clHam}
    H(p,\theta,t) = \frac{p^2}{2}+k \sum_{n}\delta\left({\frac{t}{\tau}}-n\right) V_0(\theta),
\end{gather}
where $V_0(\theta)=\cos(\theta)$, giving rise to the Chirikov map~\cite{CHIRIKOV1979263}: $\{p_{n+1} \!=\! p_n \!+\! k\sin(\theta_n) \mod 2\pi; \theta_{n+1} \!=\! \theta_n \!+\! p_{n+1} \mod 2\pi\}$, where $\theta$ is the angle of rotation, $p$ is the angular momentum, $k$ is the kicking strength. The chaoticity of the system varies with kicking strength; the phase space dynamics remain regular for values of $k\ll 1$, mixed (with both regular and chaotic orbits) for $1\lesssim k\lesssim5$, and completely ergodic (chaotic) for $k\gtrsim 5$~\cite{Greene:1979sgz,MACKAY1983283,  GARREAU201731}. The transition from integrability to chaos is evident in the diffusion of the momentum expectation value, which also occurs in its quantum analog~\cite{PhysRevLett.61.659} and has been observed experimentally~\cite{PhysRevLett.75.4598, PhysRevLett.80.4111, PhysRevE.70.056206, PhysRevLett.115.240603, PhysRevLett.117.144104}. 

The QKR is easily extended from the CKR by a canonical transformation of the position and momentum coordinates, i.e., mapping momentum in Eq.~\eqref{eq:clHam} to $p \to -i\hbar_{\rm eff}\tfrac{\partial}{\partial\theta}$, where $\hbar_{\rm eff}$ is an effective Planck's constant. Instead of a rotating rod, the QKR describes a particle which is confined to move on a ring and is subjected to a periodic potential which is turned on and off instantaneously. Here, we attach a spin-$\tfrac{1}{2}$ probe to the QKR~\cite{Scharf_1989} and compute the input-output correlations of its OTOT. This is achieved by mapping $V_0(\theta) \to \sum_{i=0}^3 V_i(\theta) \!\otimes\! \sigma_i$, where $\{\sigma_i\}_{i=0,1,2,3}$ are the 2$\times$2 identity and Pauli spin operators on the Hilbert space of the probe and $V_i$ are 2$\pi$-periodic potentials of the form $\{V_0(\theta) =\cos(\theta), \ V_j(\theta) = v_j \sin(j\theta)\}_{j=1,2,3}$. We have checked that the qualitative behavior of our results is independent of the precise choice of $V_i$. As in the classical case, this Hamiltonian becomes chaotic in the large kick regime~\cite{PhysRevA.97.063603}. In particular, for an entangling $SP$ coupling the OTOT correlations decay for a chaotic $S$.

\begin{figure*}
   \makebox[\textwidth][c]{\includegraphics[width=1\textwidth]{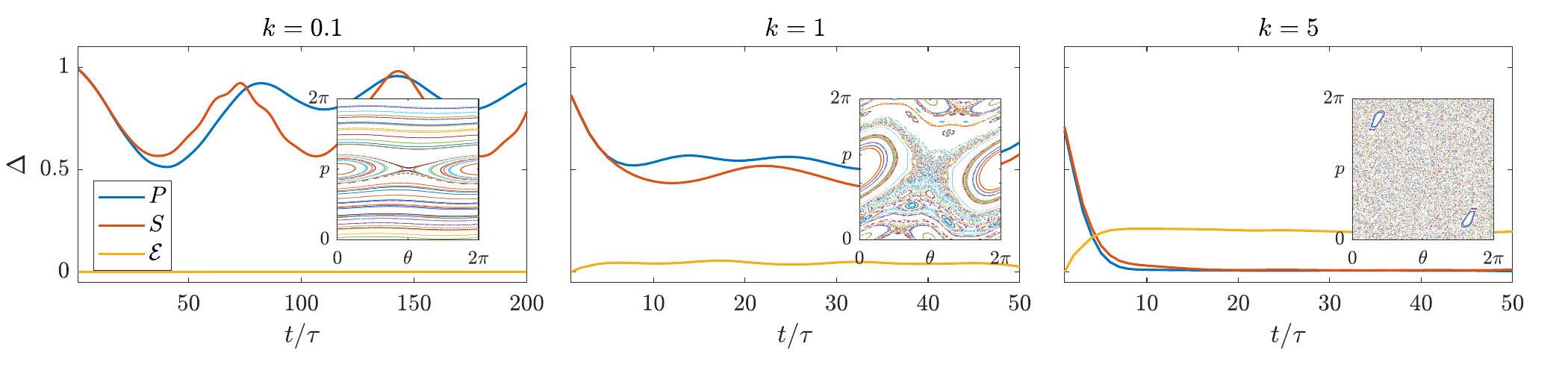}}
    \caption{The decay of $\Delta$ when the perturbation is made to the probe (blue lines) or to the system (red lines), contrasted against the Chirikov map (insets)~\cite{SM} for the corresponding kicking strength. The value of the kicking strength for the Chirikov map and for the QKR is equal in all panels and set to $0.1$, $1$ and $5$, from left to right. \add{The entanglement $\mathcal{E}$ (yellow lines) is also shown for each $k$.} The data has been smoothed using a moving average filter~\cite{SimpleMovingAverage}.}
    \label{fig:perturbations results}
\end{figure*}

In our numerics, we approximate the momentum space of the QKR by 2048 levels, and restrict $\mathcal{B}$ to a single qubit. We compute $\Delta$  with the unitary dynamics $\mathcal{U}_t$ governed by the total Hamiltonian. $k$ values are taken in the regular, intermediate and chaotic regime. The effective $\hbar$ is set to unity to reflect that we are working in the quantum regime~\cite{delande2013kicked}. Figure~\ref{fig:perturbations results} \add{(red and blue lines)} demonstrates, by direct contrast to the corresponding classical phase profiles, that OTOT correlations provide a signature of the chaoticity in the QKR. In the regular regime (left panel), the process is approximately reversible (coherent) leading to high retention of information in the OTOT, as quantified by $\Delta$ in Eq.~\eqref{eq:Delta}. Conversely, in the chaotic regime (right panel), $\Delta$ decays rapidly and saturates to zero; the OTOT becomes effectively irreversible and the quantum information that is initially stored in $P$ becomes lost in $S$ due to scrambling. Finally, in the intermediate case (middle panel), the OTOT shows a combination of both irreversible and reversible processes. In each case, a strong correspondence exists between the classical phase profiles and the correlations in OTOT. These results are in strong agreement with existing literature for OTOCs~\cite{PhysRevB.98.134303,PhysRevLett.123.160401}. \add{In addition, we show that although $\Delta$ (and consequently OTOC) decay rapidly when the QKR is chaotic, the entanglement $\mathcal{E}$ in the OTOT  (yellow line) will still grow, due to non-Markovian correlations between the forwards and backwards channels.}

We further observe in Fig.~\ref{fig:perturbations results} a remarkable similarity in the behavior of OTOT correlations regardless of whether the perturbation is applied to either $S$ or $P$ directly. This illustrates that the dynamics are sensitive to any perturbation. This makes the OTOT a particularly useful tool which can provide insight into the chaoticity of a quantum system, regardless of whether the system can be interacted with directly or not.

\textit{Conclusions.}--- We have introduced a fully quantum representation of out-of-time-order correlators as a higher order map that we have termed the out-of-time-order tensor (OTOT). The OTOT provides a number of new theoretical insights into the nature of quantum chaos and is operationally grounded and readily accessible for experimental investigations. Firstly, the information theoretic measures employed here give us a basis independent assessment of chaoticity of the process. Secondly, we demonstrate that while a quantum chaotic system may have a fast decaying OTOC, it must have slow or non-decaying non-Markovian correlations in the OTOT, which can be measured experimentally. Thirdly, we highlight the important connection between the multitime correlations in OTOT and the multitime classical correlations and classical chaos in terms of the so-called \textit{dynamical entropy} or \textit{Kolmogorov-Sinai entropy}.

On the practical side, we illustrated the utility of the OTOT as a signature for chaos in the quantum kicked rotor. Our results can be directly experimentally probed in an ultracold atomic gas. Here, the QKR has already been realized by using a pulsed standing wave lattice~\cite{Moore1995}, and it has been demonstrated that a very high resolution is achievable using a Bose-Einstein condensate~\cite{Ryu2006}. The practical realization of the QKR with spin furthermore requires the atom's momentum to be coupled to its internal state, which has recently been realized in the context of a quantum walk in momentum space~\cite{Dadras2018}.

\begin{acknowledgments}
\textbf{Acknowledgments}.--- We are grateful to Zhe-Yu Shi for early contributions to this work, and to Haydn Adlong, Neil Dowling, Kris Helmerson, and Shaun Johnstone for useful discussions. KM, JL and MMP are supported through Australian Research Council Future Fellowships FT160100073, FT160100244 and FT200100619, respectively. JL and MMP acknowledge support from the Australian Research Council Centre of Excellence in Future Low-Energy Electronics Technologies (CE170100039). KM acknowledges the support of Australian Research Council's Discovery Projects DP210100597 \& DP220101793.
\end{acknowledgments}

\bibliography{mybib}

\begin{thebibliography}{79}
\bibitem [{Sim()}]{SimpleMovingAverage}%
  \BibitemOpen
  \href@noop {} {}\bibinfo {note} {The simple moving average is given by,
  $\tilde{I}_t(n) = \frac{1}{2N+1}\sum_{i=-N}^{N}{I}_t(n-i)$ with respect to
  the true data ${I}_t(n)$, where $2N+1$ is the averaging size and we have set
  $N$=4.}\BibitemShut {Stop}%
\end{thebibliography}

\renewcommand{\theequation}{S\arabic{equation}}
\renewcommand{\thefigure}{S\arabic{figure}}
\renewcommand{\thetable}{S\arabic{table}}

\onecolumngrid

\setcounter{equation}{0}
\setcounter{figure}{0}
\setcounter{table}{0}
\setcounter{section}{0}

\clearpage

\section*{SUPPLEMENTAL MATERIAL: ``Signatures of quantum chaos in an out-of-time-order tensor''}
\setcounter{page}{1}
\begin{center}
Magdalini Zonnios, Jesper Levinsen, Meera M. Parish, Felix A. Pollock and Kavan Modi\\
\emph{\small School of Physics and Astronomy, Monash University, Victoria 3800, Australia}

\end{center}

\section{Re-writing the OTOC as an open-system map}\label{sec:OTOCtoMap}
The OTOC 
\begin{align}
F(t)=\trace\left[W(t)VW^\dag(t) V^\dag\rho\right],
\end{align}
can be written in the Schr{\"o}dinger picture as the trace over a sequence of operations made on a probe $P$ which is evolved unitarily (forwards and backwards) in time along with a system $S$, 
\begin{align}
    F(t)=&\trace\left[U_{0,t}^{SP\dag}(W^P\otimes \mathds{1}^S)U_{0,t}^{SP} (V^{P}\otimes \mathds{1}^S)U_{0,t}^{SP\dag}(W^{P\dag}\otimes \mathds{1}^S)U_{0,t}^{SP} (V^{P\dag}\otimes \mathds{1}^S)\rho^{SP}\right].
\end{align}
Superscripts denote the action of the operator either on the probe (P), system (S) or both (SP). Suppressing the identities on the system, $\mathds{1}^S$, and using the cyclicity of the trace, $\trace[ABC]=\trace[BCA]$, we have, 
\begin{align}
    F(t)=&\trace\left[U_{0,t}^{SP\dag}W^{P\dag}U_{0,t}^{SP} V^{P\dag}\rho^{SP}U_{0,t}^{SP\dag}W^PU_{0,t}^{SP} V^{P}\right].
\end{align}
Adding the implicit identities on the probe, $\mathds{1}^P$, 
\begin{align}
    F(t)=&\trace\left[\mathds{1}^P U_{0,t}^{SP\dag}W^{P\dag}U_{0,t}^{SP} V^{P\dag}\rho^{SP}\mathds{1}^P U_{0,t}^{SP\dag}W^PU_{0,t}^{SP} V^{P}\right],
\end{align}
we identify $F(t)$ as the trace over a series of maps,
\begin{align}\label{eq:F(t) trace over maps}
    F(t)&=\trace\left[\mathcal{C}^P\circ\mathcal{U}^{SP\text{ad}}_{0,t}\circ{\mathcal{B}}^{P}\circ\mathcal{U}^{SP}_{0,t}\circ \mathcal{A}^P\left[\rho^{SP}\right]\right],
\end{align}
which act as,
\begin{align}
    \mathcal{A}^{P}[\rho^P]&=V^{P\dag}\rho^P\mathds{1}^P\\
    \mathcal{U}^{SP}_{0,t}[\rho^{SP}]&={U}^{SP}_{0,t}\rho^{SP}{U}^{SP\dag}_{0,t}\\
        \mathcal{B}^{P}[\rho^P]&=W^{P\dag}\rho^P W^P\\
        \mathcal{U}^{SP\text{ad}}_{0,t}[\rho^{SP}]&={U}^{SP\dag}_{0,t}\rho^{SP}{U}^{SP}_{0,t}\\
        \mathcal{C}^{P}[\rho^P]&=\mathds{1}^P\rho^P V^{P}.
\end{align} 
We define Eq.~\eqref{eq:F(t) trace over maps} to be the out-of-time-order channel $\mathcal{O}_{t}$, 
\begin{align}
    \mathcal{O}_t[\mathcal{A},\mathcal{B},\mathcal{C}] = \trace\left[\mathcal{C}^P\circ\mathcal{U}^{ SP\text{ad}}_{0,t}\circ{\mathcal{B}}^{P}\circ\mathcal{U}^{SP}_{0,t}\circ \mathcal{A}^P\left[\rho^{SP}\right]\right].
\end{align}

\section{Matrix elements of the Floquet operator for the quantum kicked rotor}
The system-probe Hamiltonian $H_{SP}$ which describes the quantum kicked rotor along with the spin is,
\begin{align}
    H_{SP}(t) = - \frac{\hbar_{\rm eff}^2}{2} \pdv[2]{\theta}\!\otimes\!  \sigma_0 +k \sum_{n}\delta\left(t/\tau-n\right) \sum_{i=0}^3 V_i(\theta) \!\otimes\! \sigma_i
\end{align}
 where $\{\sigma_i\}_{i=0,1,2,3}$ are the 2$\times$2 identity and Pauli spin operators on the Hilbert space of the probe and $V_i$ are 2$\pi$-periodic potentials of the form $\{V_0(\theta) =\cos(\theta), \ V_j(\theta) = v_j \sin(j\theta)\}_{j=1,2,3}$. In our numerical simulations, we have set $v_1=0.1$, $v_2=0.2$, $v_3=0.3$.
Solving the (dimensionless) Schr\"odinger equation $i\hbar_{\rm eff}\pdv{(t/\tau)}U = H U$ over a single period, ${t\in(0,\tau]}$, we obtain the Floquet operator~\cite{PhysRevA.100.053608, PhysRevA.97.063603, PhysRevB.96.144301} $U_{t,t+\tau}$, which evolves a state from time $t$ to the time $t+\tau$ one period later. The periodicity of $H_{SP}(t)$ allows the action of $U_{t,t+n\tau}$ to be achieved by repeated action of the Floquet operator, $U_{t,t+n\tau} = U_{t,t+\tau}^n$, for integer $n$. Choosing the first period of the rotor, we have,
\begin{align}\label{Floquet operator}
    U_{0,\tau}&=  \lim_{\epsilon\to 0} \exp\left[-\frac{i}{\hbar_{\rm eff}} \int_{\epsilon}^{1-\epsilon}H_{SP}(t) d(t/\tau)\right]\exp\left[-\frac{i}{\hbar_{\rm eff}}    \int_{1-\epsilon}^{1+\epsilon}H_{SP}(t) d(t/\tau)\right] \\
    &= \exp\left[    \frac{i\hbar_{\rm eff}}{2}\pdv[2]{\theta}\!\otimes\!\mathds{1}^P\right]\exp\left[-ik (V_0(\theta)\!\otimes\!\mathds{1}^P+\bm{V}(\theta)\cdot\bm{\sigma}) \right],
\end{align}
or simply, 
\begin{align}\label{eq:Umm}
    U_{0,\tau}= \exp\left[-i \hbar_{\rm eff} H_F\right]\exp\left[-i(k/\hbar_{\rm eff}) H_K\right]
\end{align}
where $H_F:= -\frac{1}{2} \pdv[2]{\theta}\!\otimes\! \sigma_0$ and $H_K:= \sum_{i=0}^3 V_i(\theta) \!\otimes\! \sigma_i$.

To proceed, we note that $H_{SP}(t)$ is fixed by two (angular) momentum indices $m,m'\in[-N+1,N]$, and its elements are specified by two spin indices $s,s'\in\{0,1\}$. In our simulations we set $N=660$, which we have checked leads to fully converged numerical results. An eigenstate $\ket{m}$ of the momentum operator $p$ can be expressed in the position representation as $\braket{\theta}{m}:= e^{i\theta m}$, with corresponding eigenvalue $\hbar_{\rm eff}m$. Elements $H^{mm'ss'}:=\mel{m'}{H^{ss'}}{m}$ are thus obtained via the integral,
\begin{align}
    {H^{mm'ss'}} = \frac{1}{2\pi}\int_{0}^{2\pi}e^{i\theta(m-m')} H^{ss'}{(t)} d\theta.
\end{align}

Choosing potential terms ${V_0(\theta)=\cos{\theta}}$, ${V_1(\theta)=v_1\sin{\theta}}$, ${V_2(\theta)=v_2\sin{2\theta}}$, ${V_3(\theta)=v_3\sin{3\theta}}$ and setting ${m-m'=r}$, we have
\begin{align}\label{eq:HF els}
    {H_F^{mm'ss'}} = \frac{m^2}{2} \delta_{r,0} \delta_{s,s'}
\end{align}
and 
\begin{subequations}\label{eq:HK els}
\begin{align}
    {H^{mm'00}_K} &= \frac{1}{2}\left[ \delta_{r,1} + \delta_{r,-1} + iv_3 \delta_{r,3} - iv_3 \delta_{r,-3}\right]\\
    {H^{mm'01}_K} &= \frac{1}{2}\left[iv_1\delta_{r,1}-iv_1\delta_{r,-1}-v_2\delta_{r,-2}+v_2\delta_{r,2}\right]\\
    {H^{mm'10}_K} &= \frac{1}{2}\left[iv_1\delta_{r,1}-iv_1\delta_{r,-1}+v_2\delta_{r,-2}-v_2\delta_{r,2}\right]\\
    {H^{mm'11}_K} &= \frac{1}{2}\left[ \delta_{r,1} + \delta_{r,-1} + iv_3 \delta_{r,-3} - iv_3 \delta_{r,3} \right],
\end{align}
\end{subequations}
where we have used the property $\delta_{k,j}=\frac{1}{2\pi}\int_0^{2\pi}e^{i(k-j)\theta}d\theta$. 
The full $H_K$ and $H_F$ are obtained by summing over all $m$ and $s$ in Eq.~\eqref{eq:HF els} and Eqs.~{\eqref{eq:HK els}}, 
\begin{align}
    H_F &= \sum_{mm'ss'}H^{mm'ss'}_{F}\dyad{ms}{m's'}\\
    H_K &= \sum_{mm'ss'}H^{mm'ss'}_{K}\dyad{ms}{m's'}.
\end{align}

\section{Details of the Choi state}\label{app:choi}
The Choi state $\Upsilon^{a_\mathsf{o}c_\mathsf{i}|\mathcal{B}}$ is 
\begin{align}
\Upsilon^{a_\mathsf{o}c_\mathsf{i}|\mathcal{B}} =\frac{1}{d}\sum_{jnkm}\mathcal{O}_t\left[\dyad{jn}{km}\right]\otimes\dyad{jn}{km},    
\end{align}
where $d$ is the dimension of the probe and $\{\dyad{jn}{km}\}$ is an orthornormal basis of the superoperator space $a_\mathsf{i}\otimes a_\mathsf{o}$ and for some ancillary space of the same dimensions. The OTOT acts on the basis superoperators as,
\begin{align}
\begin{split}
    \mathcal{O}_t\left[\dyad{jn}{km}\right] = \trace_S\left[\mathcal{U}_{0,t} \circ \mathcal{B}\circ\mathcal{U}_{0,t}^\dag\left[\dyad{j}{k}\rho^{SP}_{0}\dyad{n}{m}\right]\right],
\end{split}
\end{align}
where $\rho_0^{SP}$ is the initial state of the system-probe. In our numerics we take a separable initial state $\rho_0^{SP} = \rho_0^{S}\otimes\rho_0^P$ (with $\rho_0^{S}$ in a zero momentum state and $\rho_0^P$ in a spin-up state), which effectively removes the ${a_\mathsf{i}}$ space. The Choi state satisfies,
\begin{align}
    \mathcal{O}_t[\mathcal{A}] = d\trace_{a_\mathsf{o}}[\Upsilon^{a_\mathsf{o}c_\mathsf{i}|\mathcal{B}}(\mathds{1}_{c_\mathsf{i}}\otimes{\hat\Upsilon_\mathcal{A}}^T)],
\end{align}
where subscripts the partial trace is taken over the input space, $\mathds{1}_{c_\mathsf{i}}$ is the identity operator on the output space, and $\hat\Upsilon_\mathcal{A}^T$ is the transposed Choi state of the super-operator $\mathcal{A}$.

The Choi state $\Upsilon_t^{a_\mathsf{i}c_\mathsf{o}|\mathcal{B}}$ is the state obtained by preparing the probe ($P$) in a maximally entangled state, $\sum_{jk}\frac{1}{d}\dyad{jj}{kk}$, with an ancillary system ($A$) and passing $P$ through the channel $\mathcal{O}_t$ while $A$ is acted on by the identity alone. For an initially uncorrelated system-probe state, the Choi state of $\mathcal{O}^\mathcal{B}$ is,
\begin{align}\label{eq:choirhoas}
   \Upsilon_t^{a_\mathsf{i}c_\mathsf{o}|\mathcal{B}} =\sum_{jk}\frac{1}{d} \trace_S[U_t^\dag\mathcal{B} U_t\dyad{jj}{kk}\otimes\rho_S U_t^\dag\mathcal{B}^\dag U_t],
\end{align}
where $\mathcal{B} = \cos_{\phi}\mathds{1}+i\sin_{\phi}\sigma_z$. Eq.~\eqref{eq:choirhoas} becomes, 
\begin{align}\label{eq:choirhoas2}
    \begin{split}
       \Upsilon_t^{a_\mathsf{i}c_\mathsf{o}|\mathcal{B}} =&\sum_{jk}\frac{1}{d} \trace_S[U_t^\dag(\cos_{\phi}\mathds{1}+i\sin_{\phi}\sigma_z)U_t\dyad{jj}{kk}\otimes\rho_SU_t^\dag(\cos_{\phi}\mathds{1}-i\sin_{\phi}\sigma_z)U_t]\\
        =&\sum_{jk}\frac{1}{d} \trace_S[U_t^\dag\cos_{\phi}\mathds{1}U_t\dyad{jj}{kk}\otimes\rho_SU_t^\dag\cos_{\phi}\mathds{1}U_t-i U_t^\dag\cos_{\phi}\mathds{1} U_t\dyad{jj}{kk}\otimes\rho_SU_t^\dag\sin_{\phi}\sigma_zU_t\\
        &+iU_t^\dag\sin_{\phi}\sigma_zU_t\dyad{jj}{kk}\otimes\rho_SU_t^\dag\cos_{\phi}\mathds{1}U_t+U_t^\dag\sin_{\phi}\sigma_z U_t\dyad{jj}{kk}\otimes\rho_SU_t^\dag\sin_{\phi}\sigma_zU_t]\\
        =&\sum_{jk}\frac{1}{d} \trace_S[\cos^2_\phi\dyad{jj}{kk}\otimes\rho_S-i \cos_{\phi}\sin_{\phi}\dyad{jj}{kk}\otimes\rho_S\sigma_z(t)\\
        &+i\cos_{\phi}\sin_{\phi}\sigma_z(t)\dyad{jj}{kk}\otimes\rho_S+\sin^2_\phi\sigma_z(t)\dyad{jj}{kk}\otimes\rho_S\sigma_z(t)],
        \end{split}
\end{align}
where $\sigma_z(t)=(U_t^\dag\otimes\mathds{1}^A)(\sigma_z\otimes\mathds{1}^S\otimes\mathds{1}^A)(U\otimes\mathds{1}^A)$. Re-writing the Choi state $\Upsilon^{a_\mathsf{i}c_\mathsf{o}|\mathcal{B}}_t$ in terms of an operator we define as $Z_t := U_t \sigma_z \otimes \openone U_t^\dag$ and writing the maximally mixed state $\sum_{jk}\dyad{jj}{kk}:=\psi^+$, we have,
\begin{gather}\label{eq: Choi for two-leve}
\begin{split}
        \Upsilon^{a_\mathsf{i}c_\mathsf{o}|\mathcal{B}}_t =&\cos_\phi^2 \psi^+ +\sin_\phi^2 \tr_S\left[Z_t \psi^+ \otimes \rho_S  Z_t^\dag
    + i \cos_\phi \sin_\phi \left(Z_t \psi^+ \otimes \rho_S - \psi^+ \otimes \rho_S  Z_t^\dag\right)\right].
\end{split}
\end{gather}
Removing trivial correlations by setting $\phi=\frac{\pi}{2}$, the Choi state reduces to, 
\begin{gather}
        \Upsilon^{a_\mathsf{i}c_\mathsf{o}|\mathcal{B}}_t = \tr_S[Z_t \psi^+ \otimes \rho_S  Z_t^\dag].
\end{gather}

\end{document}